%% file: eprint.tex
\documentclass[12pt]{article}
\usepackage{graphicx}
\usepackage{subfig}
\usepackage{lineno}

\newcommand\pubnumber{SNSN-323-63}
\newcommand\pubdate{\today}

\def\institute{INFN Sezione di Roma 1 \& Dipartimento di Fisica, \\
Sapienza Università di Roma, Roma; Italy}
\def\support{\footnote{Copyright 2020 CERN for the benefit of the ATLAS Collaboration. CC-BY-4.0 license.}}

\def\Title#1{\begin{center} {\Large #1 } \end{center}}
\def\Author#1{\begin{center}{ \sc #1} \end{center}}
\def\Address#1{\begin{center}{ \it #1} \end{center}}

\newcommand\pubblock{\rightline{\begin{tabular}{l} \pubnumber\\
         \pubdate  \end{tabular}}}
\newenvironment{Abstract}{\begin{quotation}  }{\end{quotation}}
\newenvironment{Presented}{\begin{quotation} \begin{center} 
             PRESENTED AT\end{center}\bigskip 
      \begin{center}\begin{large}}{\end{large}\end{center} \end{quotation}}

\input econfmacros.tex

\begin{document}
\begin{titlepage}
\pubblock

\vfill
\Title{Measurements of asymmetries in top-quark production and tests of lepton universality in ATLAS}
\vfill
\Author{ Nello Bruscino, on behalf of the ATLAS Collaboration~\support}
\Address{\institute}
\vfill
\begin{Abstract}
The top quark is the heaviest known fundamental particle. 
As it is the only quark that decays without hadronisation, it provides the unique opportunity to probe the properties of bare quarks at the Large Hadron Collider (LHC). 
This article will present two recent measurements of the top quark using $13\,\textnormal{TeV}$ collision data with the ATLAS experiment:
the top and anti-top quark pair ($t\bar{t}$) charge asymmetry measurement alongside the first test of lepton flavour universality (LFU) of leptons to $W$ bosons from $t\bar{t}$ events.
\end{Abstract}
\vfill
\begin{Presented}
$13^\mathrm{th}$ International Workshop on Top Quark Physics\\
Durham, UK (videoconference), 14--18 September, 2020
\end{Presented}
\vfill
\end{titlepage}
\def\thefootnote{\fnsymbol{footnote}}
\setcounter{footnote}{0}

\section{Introduction}
The large mass of the top quark, which is close to the electroweak symmetry breaking scale, indicates that this particle could play a special role in the Standard Model (SM) as well as in beyond the Standard Model (BSM) theories. 
Moreover, the top quark has a very short lifetime ($\tau=0.5 \times 10^{-25}\,\textnormal{s}$) and decays before hadronisation ($\tau_\textnormal{\scriptsize had} \sim 10^{-23}\,\textnormal{s}$) or spin de-correlation take place ($\tau_\textnormal{\scriptsize spin dec.} \sim 10^{-22}\,\textnormal{s}$).
Therefore several properties of the top quark may be measured precisely from its decay products.

Due to the large top-pair production ($t\bar{t}$) cross section for $13\,\textnormal{TeV}$ proton–proton ($pp$) collisions, the Large Hadron Collider (LHC) experiments collect an unprecedented number of top-quark events.
The copious amount of detected events allows for high precision measurements in order to probe predictions of quantum chromodynamics (QCD), which provides the largest contribution to $t\bar{t}$ production. 
This process may also be employed to produce a large, unbiased sample of $W$-bosons and study its properties.

This article focuses on two recent results in the top-quark sector by the ATLAS~\cite{ATLAS} Collaboration, using proton-proton ($pp$) collisions at the Large Hadron Collider (LHC):
\begin{itemize}
\item the inclusive and differential measurements of the charge asymmetry ($A_C$) in $t\bar{t}$ events at $13\,\textnormal{TeV}$;
\item the first test of lepton-flavour universality using di-leptonic $t\bar{t}$ events at $13\,\textnormal{TeV}$.
\end{itemize}

\section{Measurements of the charge asymmetry in $t\bar{t}$ events at $13\,\textnormal{TeV}$ with the ATLAS detector}
Production of top quark pairs is symmetric at leading-order (LO) under charge conjugation.
The asymmetry between the $t$ and $\bar{t}$ originates from interference of the higher-order amplitudes in the $q\bar{q}$ and $qg$ initial states, with the $q\bar{q}$ annihilation contribution dominating.
The contribution from electro-weak corrections is about 13\% for the inclusive asymmetry.
The $gq \to t\bar{t}q$ production process is also asymmetric, but its cross section is much smaller than $q\bar{q}$.
Gluon fusion production is symmetric to all orders.
As a consequence of these asymmetries, the top quark is preferentially produced in the direction of the incoming quark.

At a $p\bar{p}$ collider, where the preferential direction of the incoming quark (antiquark) always almost coincides with that of the proton (anti-proton), a forward-backward asymmetry $A_\textnormal{\scriptsize FB}$ can be measured directly.
At the LHC $pp$ collider, since the colliding beams are symmetric, it is not possible to measure $A_\textnormal{\scriptsize FB}$ as there is no preferential direction of either the top quark or the top antiquark.
However, due to the difference in the proton parton distribution functions, on average the valence quarks carry a larger fraction of the proton momentum than the sea antiquarks.
This results in more forward top quarks and more central top antiquarks.

A central-forward charge asymmetry for the $t\bar{t}$ production, referred to as the charge asymmetry ($A_\textnormal{\scriptsize C}$), is defined as $A_\textnormal{\scriptsize C}^{t\bar{t}}=\frac{N(\Delta|y|>0)-N(\Delta|y|<0)}{N(\Delta|y|>0)+N(\Delta|y|<0)}$,
where $\Delta|y|=|y(t)|-|y(\bar{t})|$ is the difference between the absolute value of the top-quark rapidity $|y_t|$ and the absolute value of the top-antiquark rapidity $|y_{\bar{t}}|$.

The measurement of the $t\bar{t}$ charge asymmetry is performed using data corresponding to an integrated luminosity of $139\,\textnormal{fb}^{-1}$ from the ATLAS experiment~\cite{bib:asymmetry}.
It is performed in the single-lepton channel combining both the resolved and boosted topologies of top quark decays.
A Bayesian unfolding procedure is used to infer the asymmetry at parton level, correcting for detector resolution and acceptance effects.

\begin{figure}[!htb]
\centering
\subfloat[]{\includegraphics[width=0.33\linewidth]{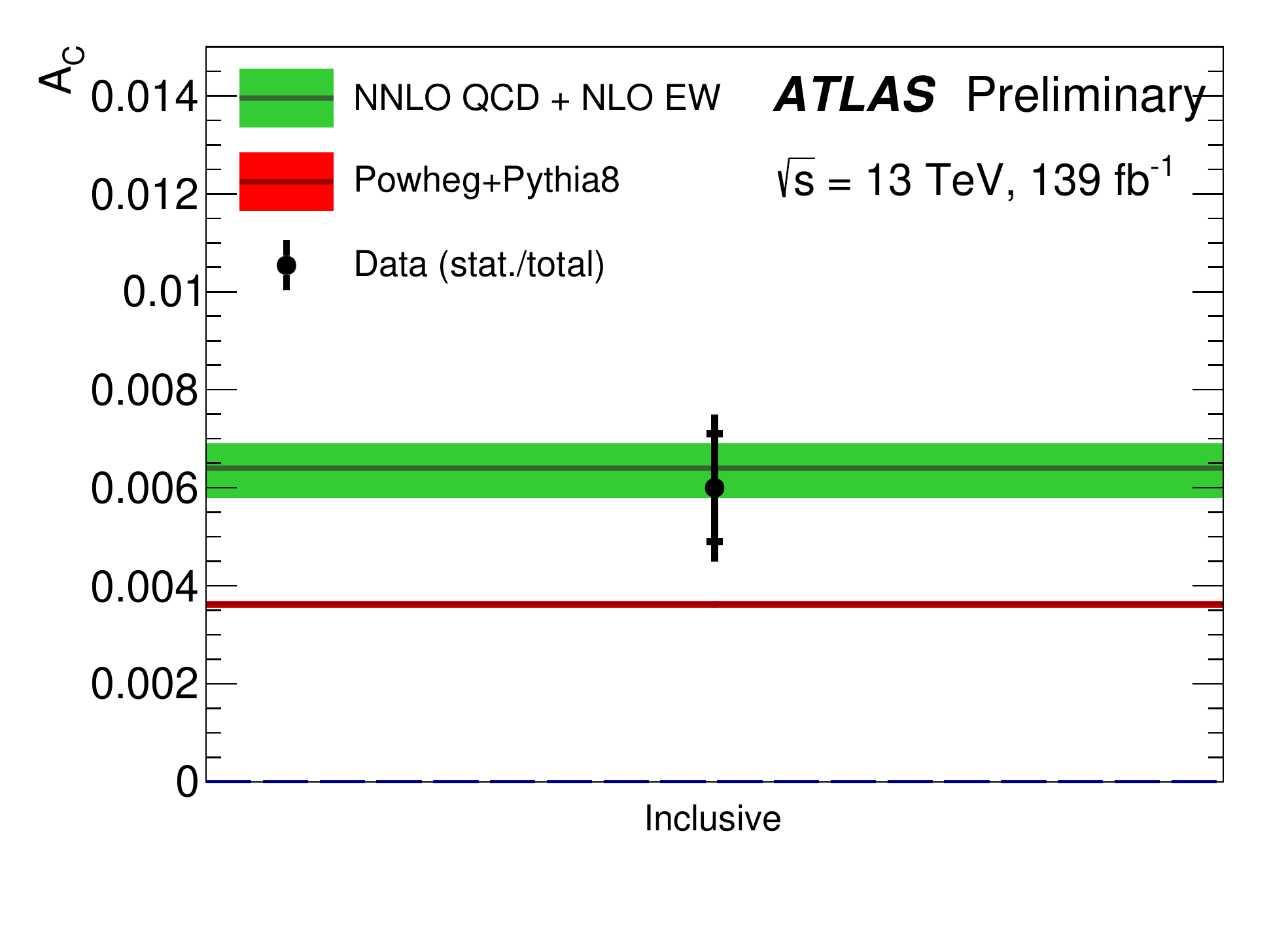}}
\subfloat[]{\includegraphics[width=0.33\linewidth]{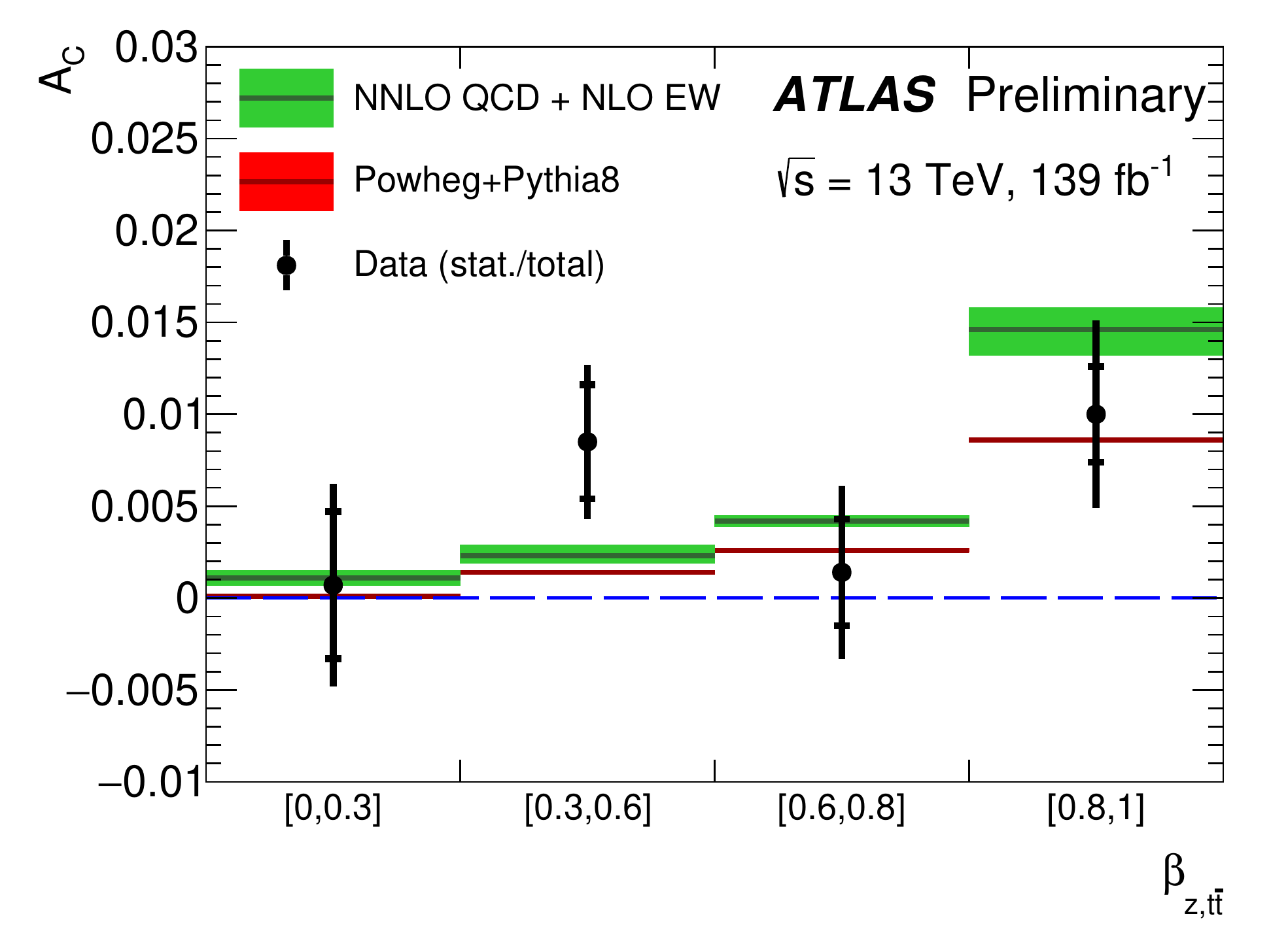}}
\subfloat[]{\includegraphics[width=0.33\linewidth]{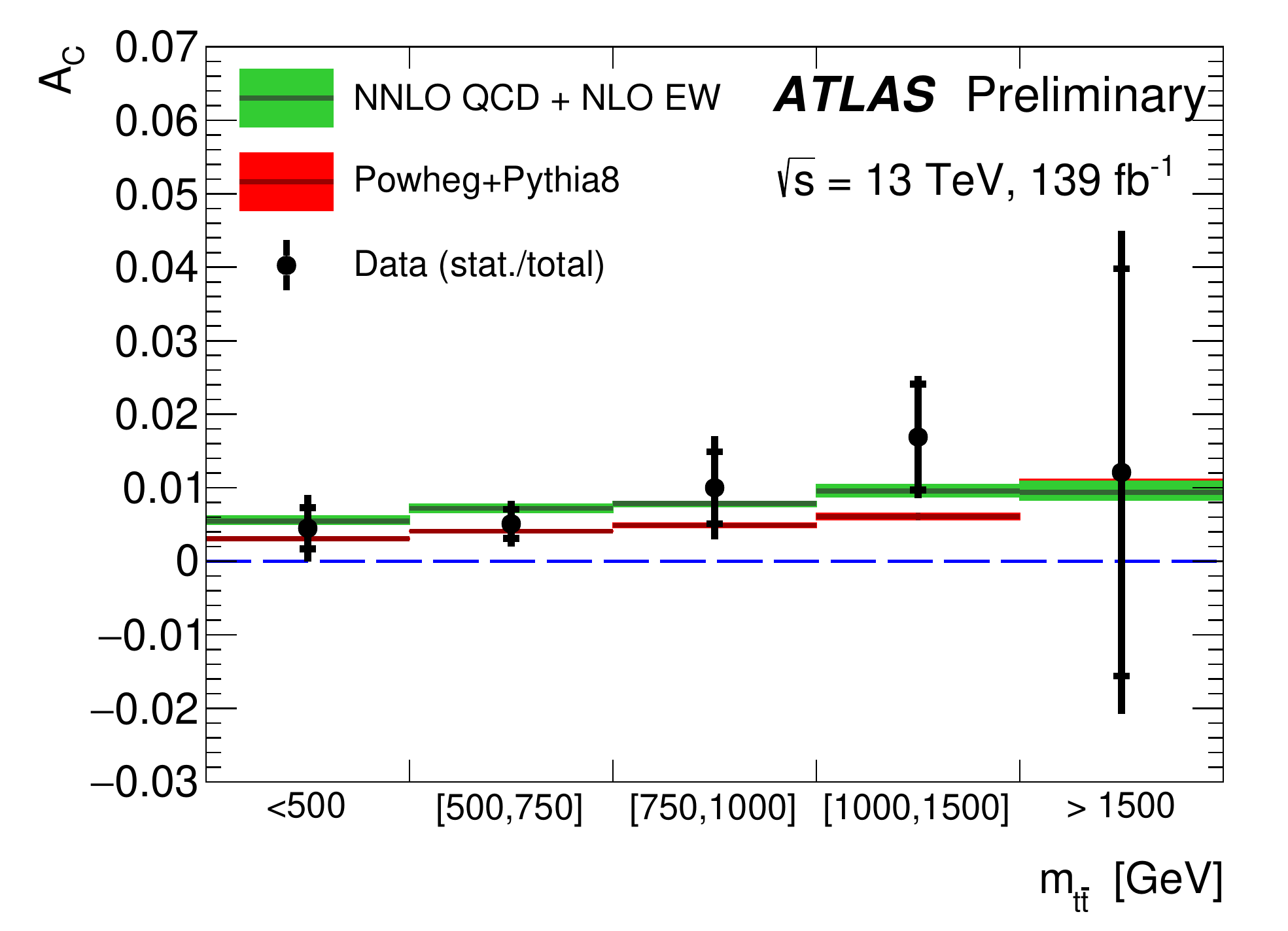}}
\caption{The unfolded inclusive (a) and differential charge asymmetries as a function of the invariant mass (b) and the longitudinal boost (c) of the top pair system in data (resolved and boosted topologies are combined). 
  Green hatched regions show SM theory predictions calculated at NNLO in QCD and NLO in electroweak theory. 
  Red hatched regions show parton-level truth asymmetry with its uncertainty extracted from the full phase space using the nominal $t\bar{t}$ signal sample. 
  Vertical bars correspond to the total uncertainties.~\cite{bib:asymmetry}}
\label{fig:prop_asymmetry}
\end{figure}

The inclusive $t\bar{t}$ charge asymmetry is measured as $A_\textnormal{\scriptsize C} = 0.0060 \pm 0.0015 \textnormal{(stat+syst.)}$, which differs from zero by 4 standard deviations.
It corresponds to the first evidence for charge asymmetry in $pp$ collisions.
Differential measurements are performed as a function of the invariant mass and longitudinal boost of the $t\bar{t}$ system.
Both inclusive and differential measurements are found to be compatible with the SM predictions, at NNLO in perturbation theory with NLO electroweak corrections,
and are shown in Figure~\ref{fig:prop_asymmetry}.

\section{Test of the universality of $\tau$ and $\mu$ lepton couplings in $W$-boson decays from $t\bar{t}$ events with the ATLAS detector}
The Standard Model of particle physics encapsulates our current best understanding of physics at the smallest scales. 
A fundamental axiom of this theory is the universality of the couplings of the different generations of leptons to the electroweak gauge bosons. 
The measurement of the ratio of the rate of decay of $W$ bosons to $\tau$-leptons and muons, $R(\tau/\mu) = B(W \to \tau\nu_\tau)/B(W \to \mu\nu_\mu)$, constitutes an important test of this axiom. 
Previously, $R(\tau/\mu)$ has been measured by the four experiments at the Large Electron–Positron Collider (LEP), yielding a combined value of $1.070 \pm 0.026$~\cite{bib:LEPlfu}. 
This deviates from the SM expectation of unity by $2.7\sigma$, motivating a precise measurement of this ratio at the LHC.

The measurement of this quantity is performed with a novel technique using the large number of $t\bar{t}$ events produced at LHC.  
It is based on $139\,\textnormal{fb}^{-1}$ of data recorded with the ATLAS detector in proton–proton collisions at $13\,\textnormal{TeV}$~\cite{bib:ttbarlfu}. 
Given the large $B(t \to W q)$, close to 100\%, a very large sample of $W$ boson pairs can be exploited. 
These are used in a tag-and-probe technique to obtain a large sample of clean and unbiased W boson decays to muons and $\tau$-leptons. 
The $\tau$-leptons are identified through their decay to muons. 
The displacement of the $\tau$ decay vertex and the different muon transverse momentum ($p_\textnormal{\scriptsize T}$) spectra are used to distinguish between muons from the $W \to \tau\nu_\tau$ and $W \to \mu\nu_\mu$ processes, to extract $R(\tau/\mu)$. 
This is achieved by utilising the precise reconstruction of muon tracks obtainable by the ATLAS experiment.

Muons originating from $W$ bosons and those originating from an intermediate $\tau$-lepton are distinguished using the lifetime of the $\tau$-lepton, through the muon transverse impact parameter, and differences in the muon transverse momentum spectra. 
The two largest backgrounds are $Z(\to \mu\mu)$+jets and events in which the probe muon does not originate from a $W$ boson decay. 
Three dedicated control regions are used to extract the normalisation of these backgrounds.

A profile likelihood fit is performed in three bins in $p_{\textnormal{\scriptsize T}\,\mu}$ (boundaries of: 5, 10, 20, $250\,\textnormal{GeV}$) and eight bins in the transverse impact parameter, $|d_{0\,\mu}|$ (boundaries of: 0, 0.01, 0.02, 0.03, 0.04, 0.06, 0.09, 0.15, $0.5\,\textnormal{mm}$), 
of the probe muon for each channel ($e–\mu$ and $\mu-\mu$), making 48 bins in total.

\begin{figure}[!htb]
\centering
\subfloat[]{\includegraphics[width=0.33\linewidth]{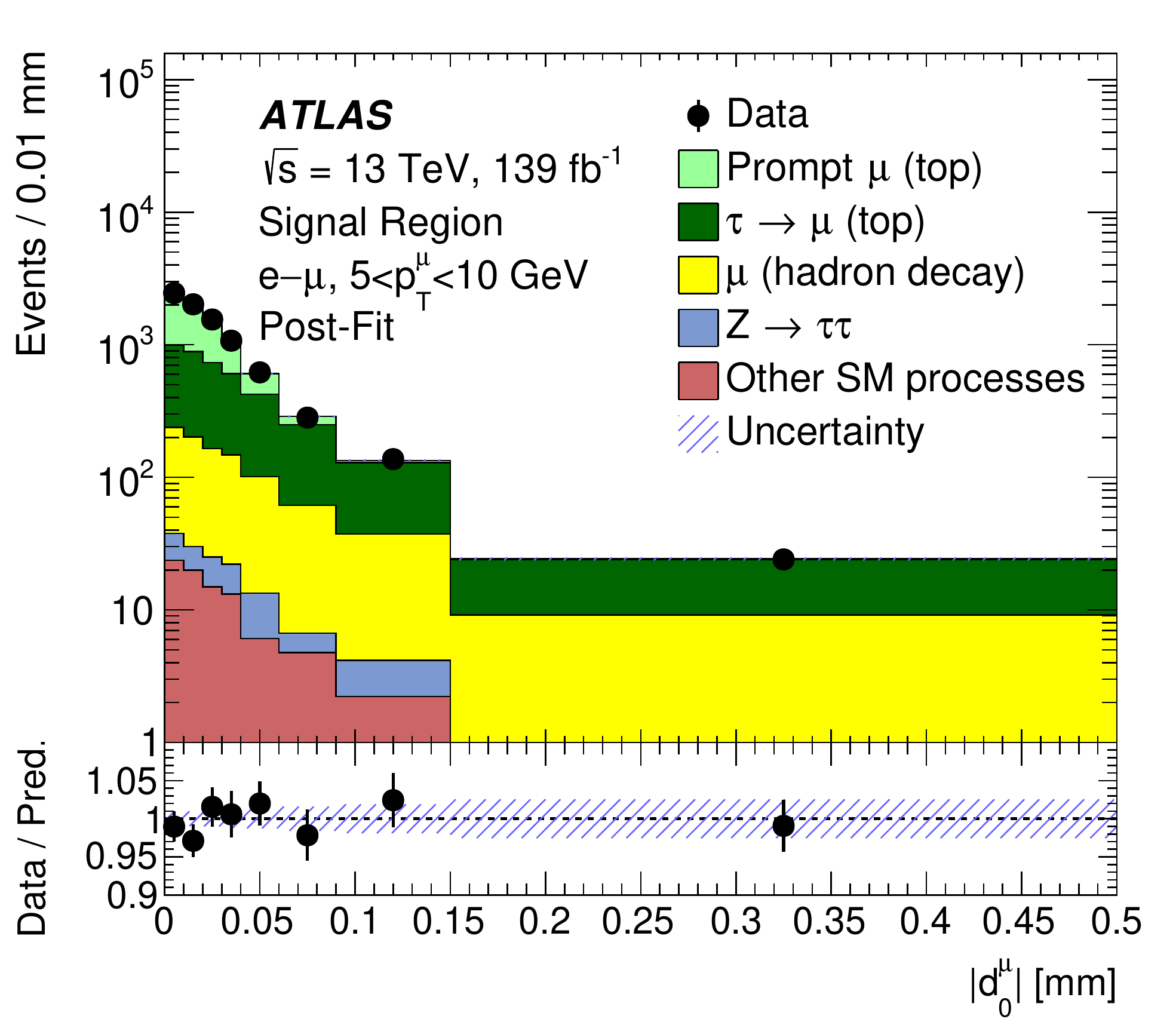}}
\subfloat[]{\includegraphics[width=0.33\linewidth]{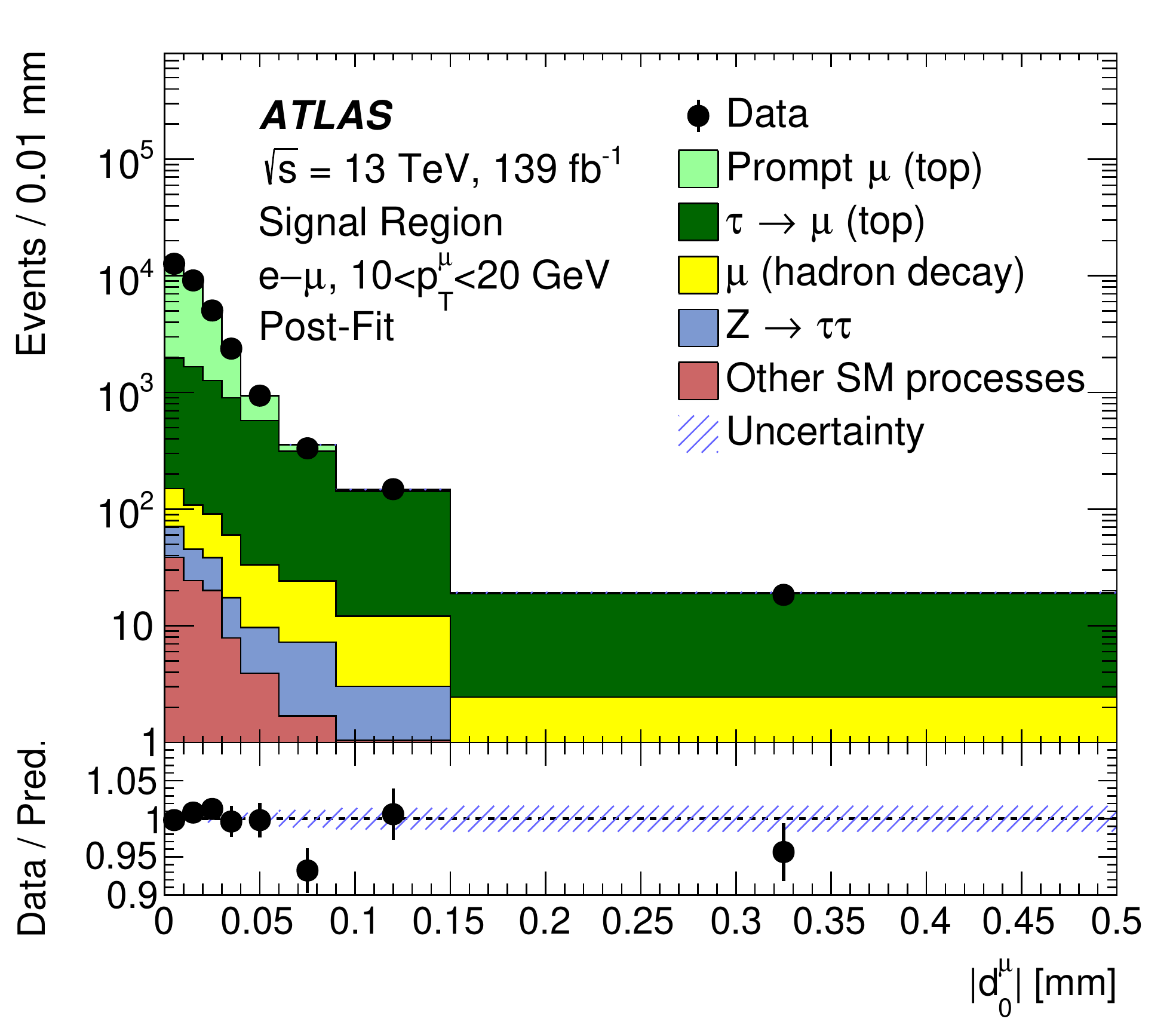}}
\subfloat[]{\includegraphics[width=0.33\linewidth]{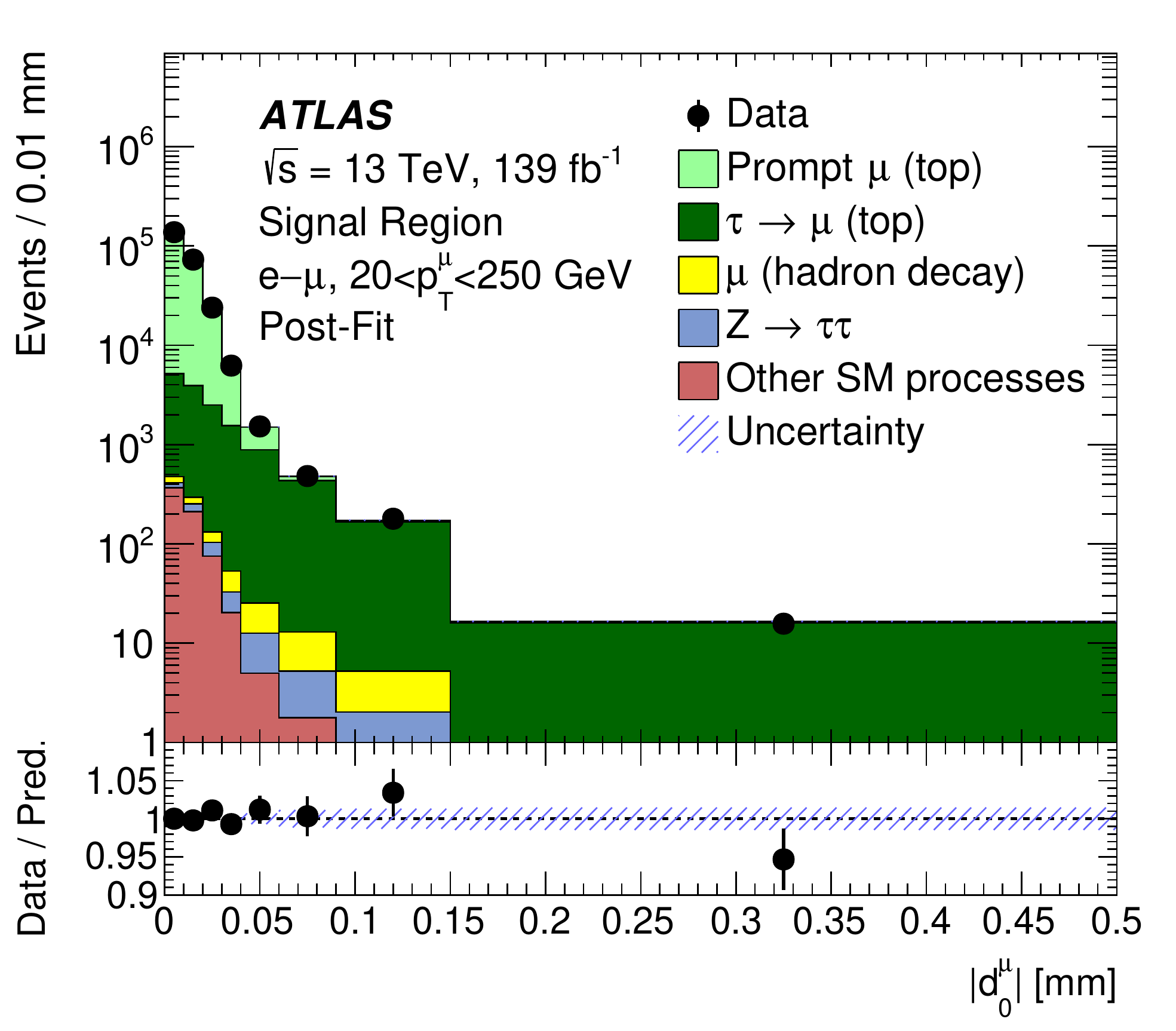}}\\
\subfloat[]{\includegraphics[width=0.33\linewidth]{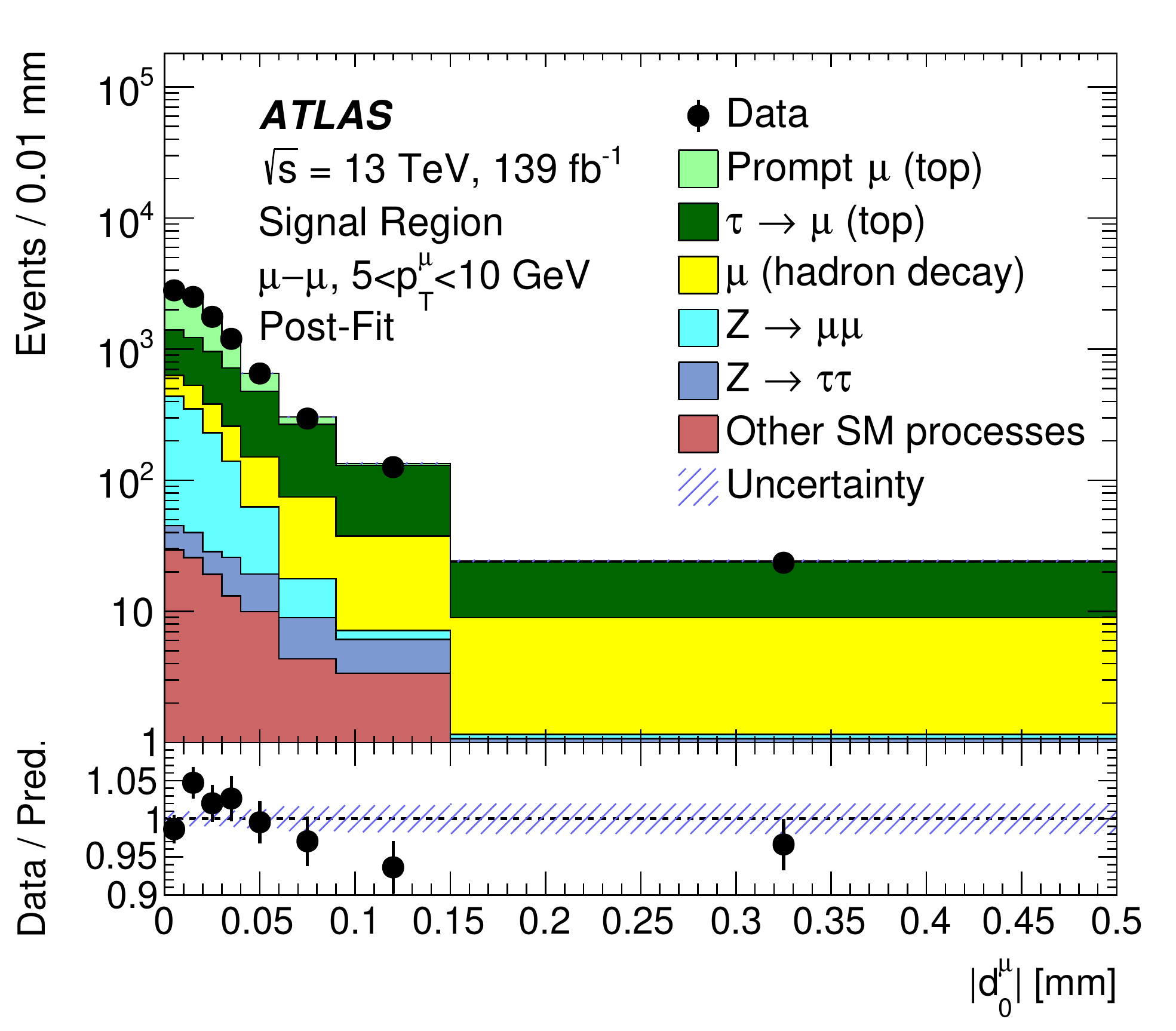}}
\subfloat[]{\includegraphics[width=0.33\linewidth]{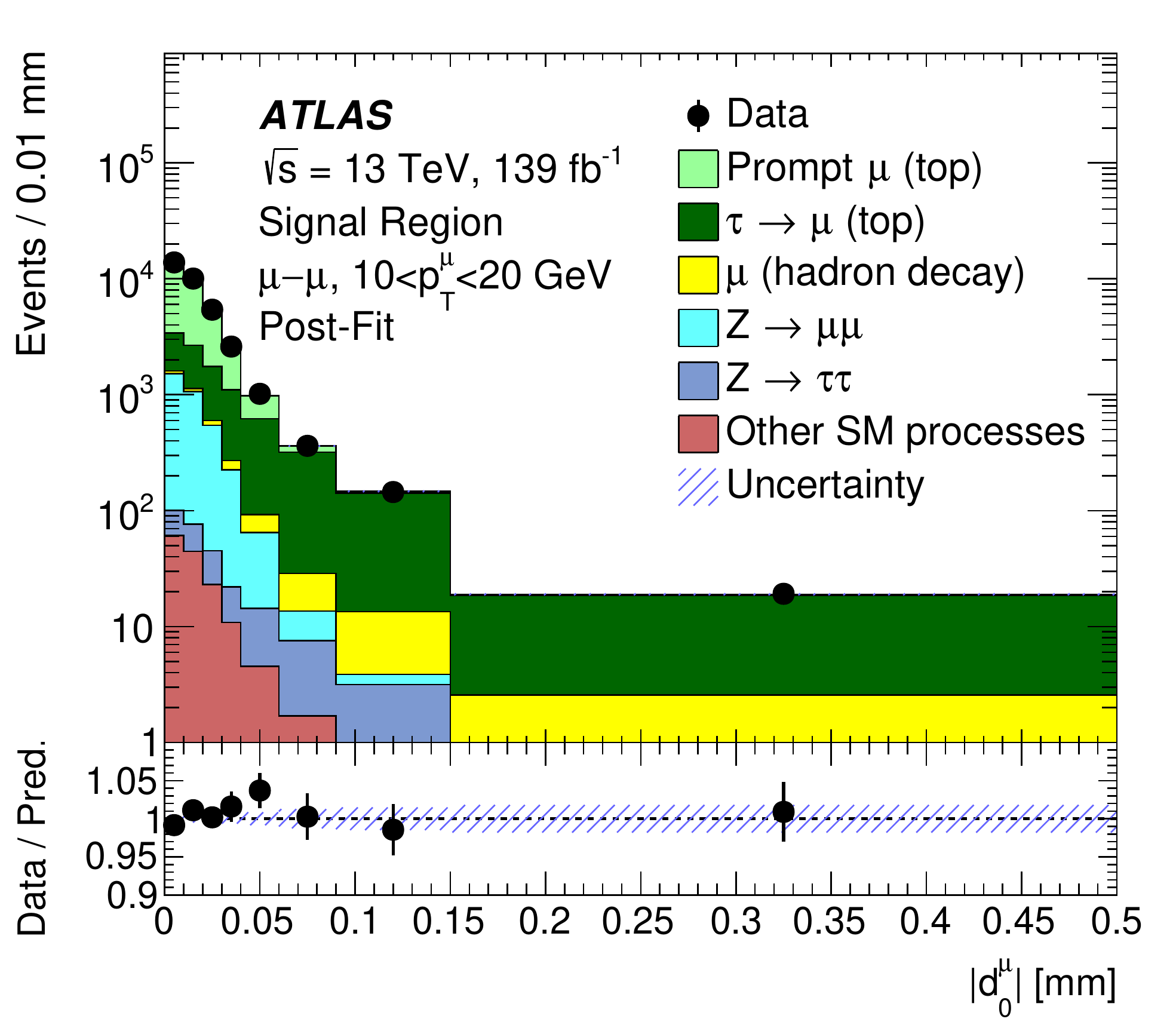}}
\subfloat[]{\includegraphics[width=0.33\linewidth]{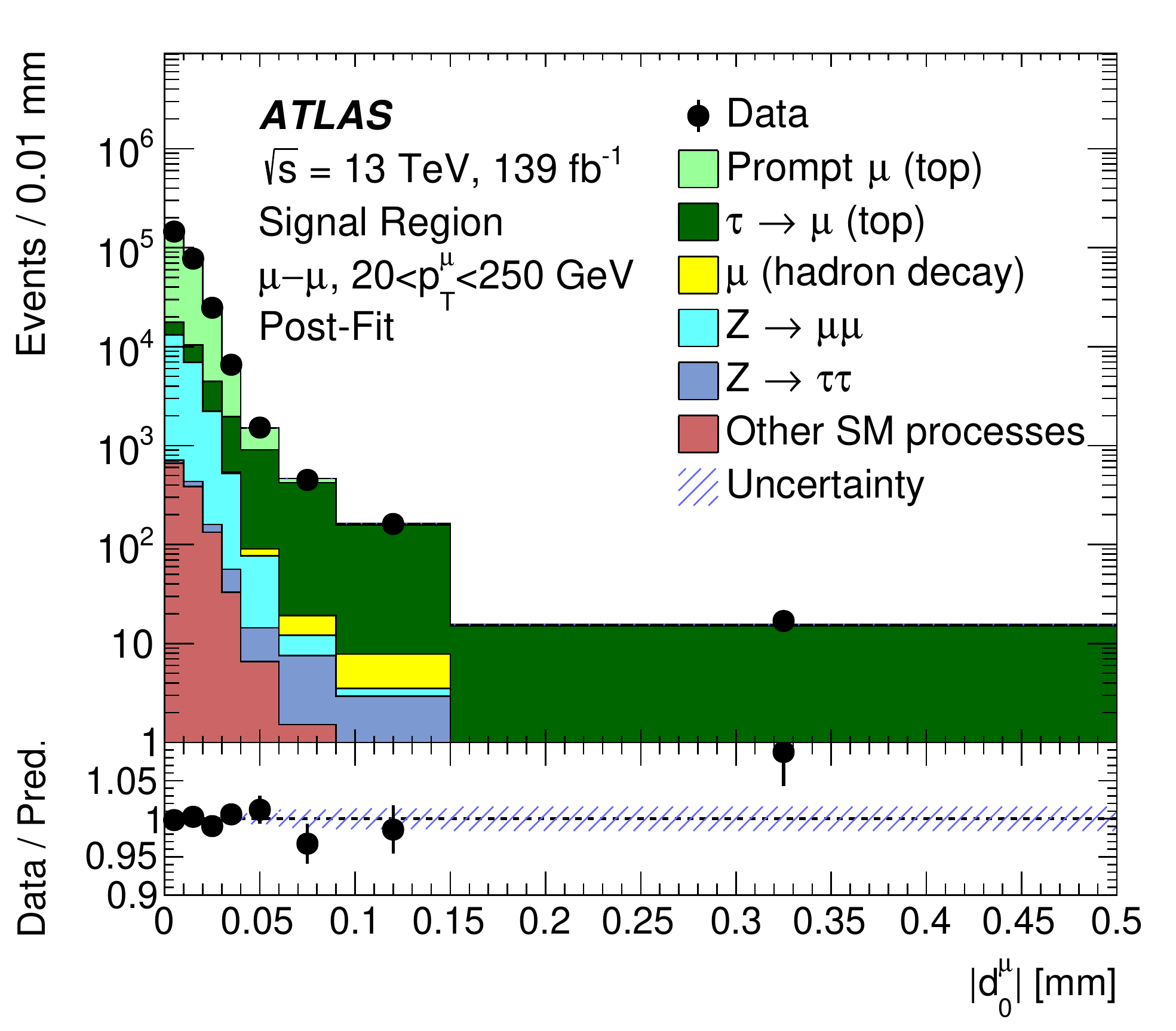}}
\caption{The $|d_{0\,\mu}|$ distributions for each channel ((a),(b),(c): $e–\mu$ channel, (d),(e),(f): $\mu-\mu$ channel) and probe muon $p_{\textnormal{\scriptsize T}\,\mu}$ bin ((a)-(d): 
  $5<p_{\textnormal{\scriptsize T}\,\mu}<10\,\textnormal{GeV}$, (b)-(e): $10<p_{\textnormal{\scriptsize T}\,\mu}<20\,\textnormal{GeV}$, (c)-(f): 2$20<p_{\textnormal{\scriptsize T}\,\mu}<250\,\textnormal{GeV}$) used in the analysis. 
  Plots are shown after the fit has been performed. 
  The data are represented by points and a stacked histogram represents the different simulated processes. 
  The bottom panel shows the ratio of the data to the expectation. 
  Blue bands indicate the systematic uncertainties with the constraints from the analysis fit applied. 
  Different components are labelled according to the muon source and process. The contribution from 'other SM processes' is dominated by di-boson and $t\bar{t}+V$ production. 
  ~\cite{bib:ttbarlfu}}
\label{fig:lfu}
\end{figure}

Figure~\ref{fig:lfu} shows the differential distributions of $|d_{0\,\mu}|$ in the six signal regions for the data and the expectation after the fit to data. 
Good agreement is observed between the corrected simulation samples and the data. 

\begin{figure}[!htb]
\centering
\includegraphics[width=0.80\linewidth]{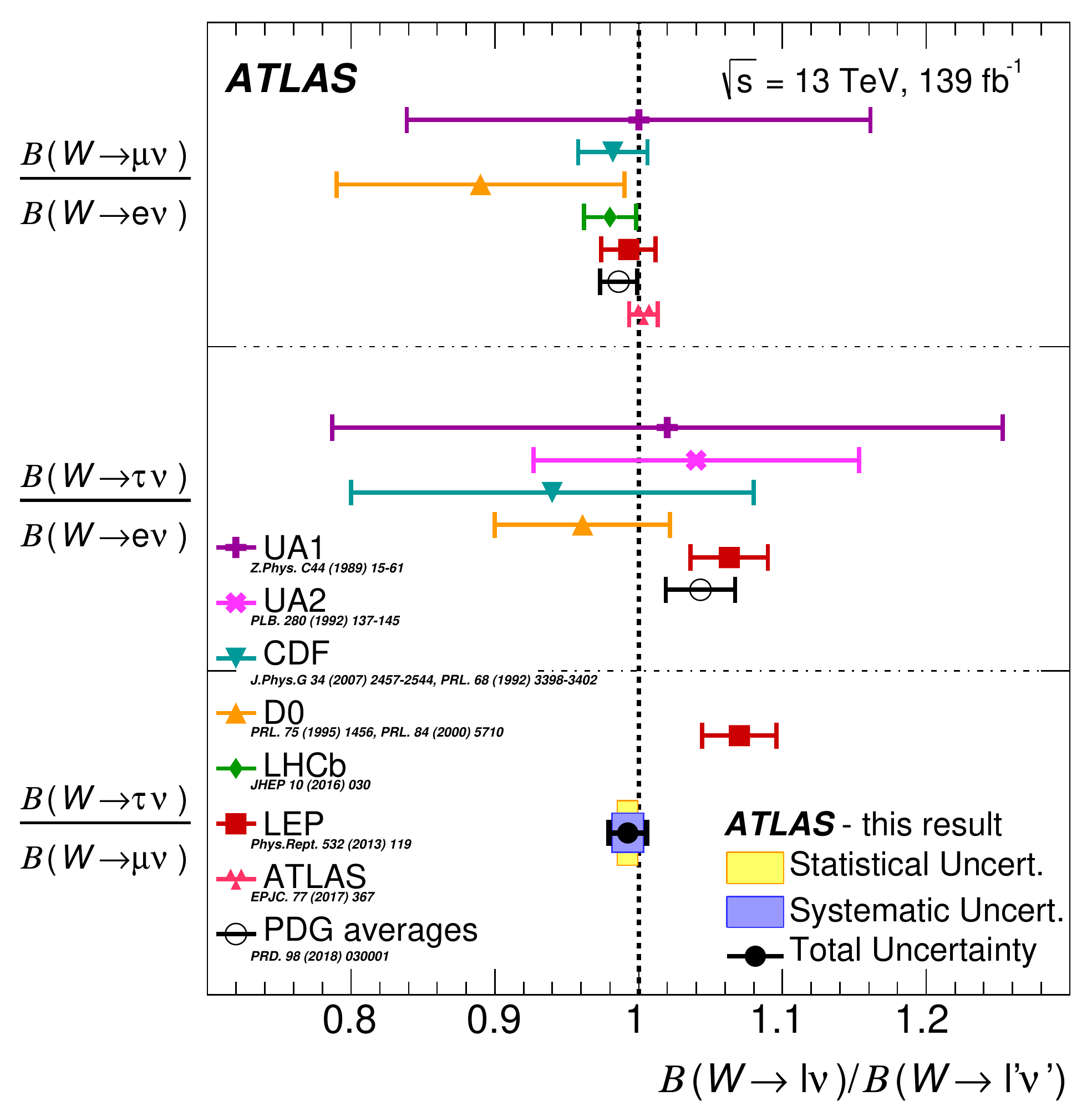}
\caption{The measured value of $R(\tau/\mu)$ with previous measurements, as well as previous measurements of $R(\mu/e)$  and $R(\tau/e)$. 
  The statistical (yellow box) and systematic (purple box) errors are shown separately as well as the total error on the measurement (black circular marker). 
  A vertical dashed line indicates the Standard Model prediction of equal branching ratios to different lepton flavours.
  ~\cite{bib:ttbarlfu}}
\label{fig:lfu_summary}
\end{figure}

The value of $R(\tau/\mu)$ is found to be $0.992 \pm 0.013$ [$\pm0.007$ (stat) $\pm 0.011$ (syst)] and is in agreement with the hypothesis of universal lepton couplings as postulated in the Standard Model. 
This is the most precise measurement of this ratio, and the only such measurement from the Large Hadron Collider, to date (Figure~\ref{fig:lfu_summary}).

\end{document}

%% file: econfmacros.tex
\def\beq{\begin{equation}}
\def\eeq#1{\label{#1}\end{equation}}
\def\eeqn{\end{equation}}

\def\beqa{\begin{eqnarray}}
\def\eeqa#1{\label{#1}\end{eqnarray}}
\def\eeqan{\end{eqnarray}}

\let\bar=\overbar

\def\Dslash{\not{\hbox{\kern-4pt $D$}}}
\def\dslash{\not{\hbox{\kern-2pt $\del$}}}

\def\msb{{\bar{\ssstyle M \kern -1pt S}}}

